\documentclass[aps,prl,preprint]{revtex4}
\usepackage{amsfonts}
\usepackage{amsmath}
\usepackage{graphicx}
\usepackage{dsfont}
\usepackage{diagbox}
\usepackage{array}
\usepackage{amssymb}
\usepackage{bbm}
\usepackage{float}
\usepackage{subfigure}
\usepackage{url}
\usepackage{array}
\usepackage{multirow}
\usepackage{bm}
\usepackage{booktabs}
\usepackage{bbm}
\usepackage{array}

\begin{document}

\title{Non-Hermitian effects of the intrinsic signs in topologically ordered
wavefunctions}
\author{Qi Zhang$^{1}$, Wen-Tao Xu$^{1}$, Zi-Qi Wang$^{1}$ and Guang-Ming
Zhang$^{1,2}$}
\affiliation{$^{1}$State Key Laboratory of Low-Dimensional Quantum Physics and Department
of Physics, Tsinghua University, Beijing 100084, China. \\
$^{2}$Frontier Science Center for Quantum Information, Beijing 100084, China.}
\date{\today}

\begin{abstract}
Negative signs in many-body wavefunctions play an important role in quantum
mechanics. The ground-state wavefunction of double semion model on a two-dimensional hexagonal lattice contains an intrinsic sign which cannot be removed by any local transformation. Here we proposed a generic double semion wavefunction in tensor network representation, and the wavefunction norm is mapped to the partition
function of a triangular lattice Ashkin-Teller model with imaginary magnetic
fields and imaginary three-spin triangular face interactions. To solve this
non-Hermitian model with parity-time (PT) symmetry, numerical tensor-network
methods are employed, and a global phase diagram is determined. Adjacent to
the double semion phase, we find a gapless dense loop phase described by
non-unitary conformal field theory and a PT-symmetry breaking phase with
zeros of the partition function. So a connection has established between the
intrinsic signs in the topologically ordered wavefunction and the PT-symmetric non-Hermitian statistical model.
\end{abstract}

\maketitle

\newpage

In quantum mechanics, negative signs in the many-body wavefunctions play an
incredibly important role, as the basic phenomenon of interference relies on
a cancellation between amplitudes of opposite sign of the wavefunctions.
Recently, topological phases of matter with anyonic excitations have
attracted considerable attention\cite%
{kitaev_TC,Freedman2003,Wen_Top,Kitaev2006,Nayak2008}. A prototype
topologically ordered state is the toric code (TC) model\cite{kitaev_TC},
and its ground-state wavefunction is an equal weight superposition of all
closed domain-wall loops. A closely related topologically ordered state is
the double semion (DS) model, whose ground-state wavefunction is a
superposition of all closed domain-wall loops weighted by a $\pi $ phase
multiplying the number of closed loops\cite{Wen_Top,LevinGu_GaugeSPT}. Due
to the presence of such non-local phase factors, low-energy excitations of
the DS phase contain semions with opposite chiralities and bosonic bound
states of semions. Actually it has been speculated that the negative signs
in the DS wavefunction are intrinsic and cannot be removed by any local
transformation\cite{Hastings2016,Freedman2016}. However, the norm of the DS
wavefunction with diagonal deformation is identical to that of TC\cite%
{HuangWei1,XuZhang2018}. Therefore, important open questions arise what
physical consequences are caused by these intrinsic signs and whether novel
anyon condensed phases are resulted in adjacent to the DS phase.

In this work, we shall resolve these questions and provide new insight into
the nature of the intrinsic signs in the topological wave functions. By
proposing a generic DS wavefunction with two distinct tuning parameters on a
two-dimensional hexagonal lattice, we explore the possible topological phase
transitions out of the DS phase. In the tensor network representation, the
non-local phase factors in the DS wavefunction can be expressed in terms of
auxiliary spins on the dual triangular lattice. By integrating out the
physical degrees of freedom, the norm of this wavefunction is mapped to the
partition function of a two-dimensional triangular lattice Ashkin-Teller
model with \textit{imaginary} magnetic fields and \textit{imaginary}
three-spin triangular face interactions, which vanish for the generic TC
wavefunction. Such a statistical model exhibits a parity-time (PT)
symmetry with negative Boltzmann weights, which are typical
features of non-Hermitian systems\cite{Bender1998,Konotop2016}. An exotic
universality class of criticality may occur due to the presence of
negative/complex Boltzmann weights, e.g., the Yang-Lee edge singularity\cite%
{Fisher1978,Cardy1985}, where an imaginary magnetic field in the
high-temperature Ising model was demonstrated to trigger an exotic phase
transition described by non-unitary conformal field theory (CFT).

To solve this PT-symmetric statistical model with negative Boltzmann
weights, we employ the numerical tensor-network methods: the
corner-transfer-matrix renormalization group (CTMRG) \cite%
{CTM_Nishino,CTM_Vidal,CTM_Corboz,fishman_faster_2018}. A global phase
diagram is fully determined in the thermodynamic limit. Adjacent to the DS
phase, we find the gapped dilute loop phase with condensed bosonic anyons,
gapless dense loop phase described by a non-unitary CFT, and a PT-symmetry breaking (PTSB) phase with zeros of the partition function. The continuous PT-symmetry breaking transitions are described by exceptional points (EPs)\cite{LeeChan2014,Ashida2017}, which have no Hermitian critical counterparts. So the derived phase diagram for the DS phase is \textit{distinctly} different from that for the TC phase. Although no dissipation is involved, the intrinsic sign problem in the topological wavefunction is closely connected with a PT-symmetric statistical model with an effective non-Hermitian Hamiltonian.

\textbf{RESULTS}

\begin{figure}[tbp]
\centering
\includegraphics[width=0.8\textwidth]{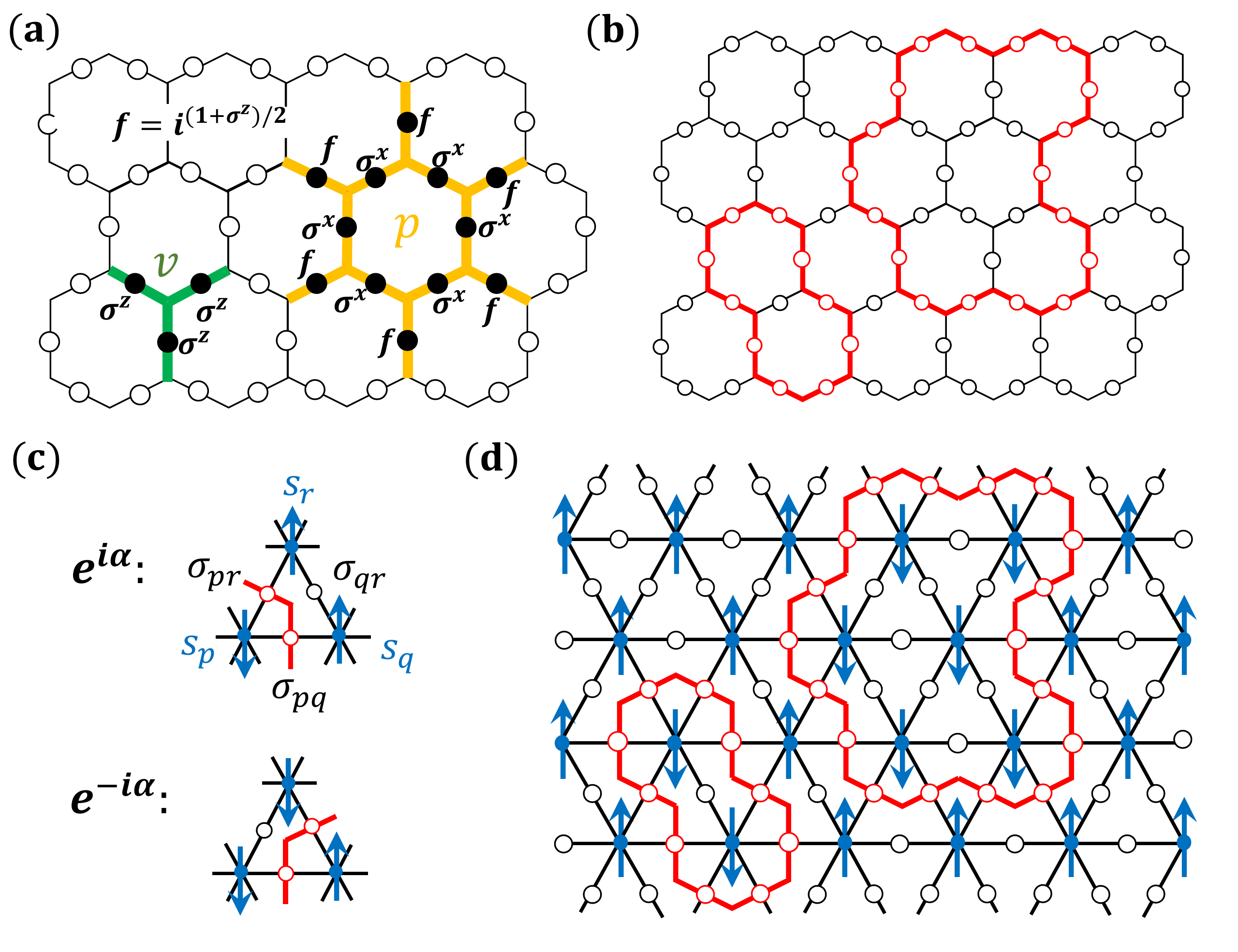}
\caption{\textbf{Definition of the double semion model and its loop representation} (a) The vertex term $A_{v}$ and plaquette term $B_{p}$ of the DS
model. (b) A typical loop configuration where the $\protect\sigma ^{z}=-1$
states form closed loops. (c) The domain-walls are assigned orientations
according to the rule that down (up) spins are on the left (right) of
orientated domain walls. A left-turn (right-turn) of domain wall is associted
with a phase factor $e^{\pm i\protect\alpha }$. (d) The auxiliary spin
configuration corresponds to the loop configuration. }
\label{parentH}
\end{figure}

\textbf{DS tensor-network wavefunction.} The DS model is defined by the
following Hamiltonian which acts on quantum spin-1/2 operators living on the
edges of a two-dimensional hexagonal lattice\cite{Wen_Top}
\begin{equation}
H_{DS}=-\sum_{v}A_{v}-\sum_{p}B_{p}.
\end{equation}%
Here the vertex term $A_{v}$ and the plaquette term $B_{p}$ shown in Fig. %
\ref{parentH}a are given by
\begin{equation}
A_{v}=\prod_{k\in E(v)}\sigma _{k}^{z},\text{ }B_{p}=\prod_{k\in E(p)}\sigma
_{k}^{x}\prod_{j\in \widetilde{E}(p)}i^{\left( 1+\sigma _{j}^{z}\right) /2},
\end{equation}%
where $E(v)$ is the set of edges around a vertex $v$, $E(p)$ is the set of
inner edges and $\widetilde{E}(p)$ is the set of outer edges around a
hexagon $p$. Operator $A_{v}=-1$ is associated to a semion or anti-semion
excitation, while $B_{p}=-1$ is a bosonic excitation of a semion-antisemion
pair. When all vertex terms $A_{v}=+1$ restrict the Hilbert space of states
to the zero-flux subspace, the ground state of the Hamiltonian is stabilized
by $B_{p}=1$ for all plaquettes:
\begin{equation}
|\Psi _{0}\rangle =\prod_{p}\frac{(1+B_{p})}{\sqrt{2}}|\uparrow \rangle
^{\otimes N},
\end{equation}%
where $|\uparrow \rangle $ is the eigenvector of $\sigma ^{z}=+1$ and $N$ is
the total number of spins. Actually, the loop representation provides a very
intuitive way to understand this wavefunction, in which $\sigma ^{z}=-1$ and
$\sigma ^{z}=+1$ states are interpreted as the presence or absence of loops
on the edges, as shown in Fig. \ref{parentH}b. Expanding this product, we
have the ground state as a superposition of all closed loop configurations
with alternative signs\cite{Wen_Top,LevinGu_GaugeSPT},
\begin{equation}
|\Psi _{0}\rangle =\sum_{\{\mathcal{L\}}}(-1)^{\#\mathcal{L}}|\mathcal{L}%
\rangle ,  \label{EqDS}
\end{equation}%
where $\#\mathcal{L}$ is the number of contractible closed loops in the
configuration $\mathcal{L}$.

Moreover, it is very useful to express the DS wavefunction in terms of
tensor-network state (TNS)\cite{gu_tensor-product_2009}. The loops can be
regarded as the domain-walls of local auxiliary Ising spins, which form a
dual triangular lattice, as shown in Fig. \ref{parentH}d. These auxiliary
Ising spins contain the crucial information about anyonic excitations\cite%
{Shadowofanyons_2015,ZhuGuoYi_TC}. Importantly, when the domain-walls of the
Ising paramagnet are assigned orientations according to the rule displayed
in Fig. \ref{parentH}c, the non-local signs in the wavefunction can be
locally expressed in terms of these auxiliary Ising spins. In the hexagonal
lattice, the difference between the left- and right-turnings for a closed
contractible orientated loop must be six, so we have $e^{i6\alpha }=-1$.
Then the many-body entangled state between the physical spins and auxiliary
spins is given by
\begin{equation}
|\tilde{\Psi}\rangle =\prod_{\langle pqr\rangle }e^{i\frac{\alpha }{4}%
(s_{p}+s_{q}+s_{r}-3s_{p}s_{q}s_{r})}\prod_{\langle pq\rangle }\frac{%
1+s_{p}s_{q}\sigma _{p,q}^{z}}{2}|s\rangle |\sigma _{p,q}^{z}\rangle ,
\label{TNS}
\end{equation}%
where $\langle pq\rangle $ denotes the nearest-neighbor dual lattice sites, $%
\langle pqr\rangle $ stands for the minimal triangular faces. The DS
tensor-network wavefunction is obtained by summing over all the auxiliary
spins: $|{\Psi }_{0}\rangle =\sum_{\{s\}}\langle s|\tilde{\Psi}\rangle $.

The above DS wavefunction is just the fixed-point wavefunction with zero
correlation length. To study the phase transitions out of the DS phase, we need a generically deformed DS wavefunction with tuning parameters, and we can employ the
wavefunction approach to reveal its essential physics. When subjected to two
different magnetic fields $h_{x}^{\prime }$ and $h_{z}^{\prime }$, the DS
model with the additional Zeeman terms are no longer exactly solvable. If
these additional terms are treated perturbably, the ground-state
wavefunction is obtained
\begin{equation}
|\Psi \left( h_{x}^{\prime },h_{z}^{\prime }\right) \rangle =\left[ 1+\frac{1%
}{8}\sum_{\langle pq\rangle }\left( h_{x}^{\prime }\sigma
_{p,q}^{x}+2h_{z}^{\prime }\sigma _{p,q}^{z}\right) \right] |\Psi
_{0}\rangle .
\end{equation}%
When the wavefunction corrections are expressed as the operator product with
the modified magnetic field parameters, we have
\begin{equation}
|\Psi (h_{x},h_{z})\rangle =\prod_{\langle pq\rangle }\left[ 1+\left(
h_{x}\sigma _{p,q}^{x}+h_{z}\sigma _{p,q}^{z}\right) \right] |\Psi
_{0}\rangle ,
\end{equation}%
which can be regarded as a generic DS wavefunction in an expanded parameter
region. Actually, the similar deformation has been used to express the
generic TC wavefunction\cite%
{PhysRevX.5.011024,PhysRevLett.111.090501,ZhuGuoYi_TC} and Fibonacci
quantum-net wavefunction\cite{FibonacciXuWenTao}. For convenience, we define
$h_{x}\equiv h\cos \theta $ and $h_{z}\equiv h\sin \theta $. When $%
h\rightarrow 1$, the deformation filters out the spin-polarized trivial
state. It should be emphasized that this generic wavefunction still has a
local parent Hamiltonian. The possible continuous quantum phase transitions
of such a Hamiltonian are characterized by the so-called conformal quantum
critical theories\cite{ardonne,castelnovo}, where all equal-time correlators
of local operators are described by two-dimensional CFT.

\textbf{Mapping to PT-symmetric statistical model.} To study the possible
topological phase transitions out of the DS phase, the norm of the deformed
DS wavefunction is considered. By summing over the physical degrees of
freedom, a double-layer tensor network is obtained
\begin{equation}
\langle \Psi (h,\theta )|\Psi (h,\theta )\rangle =\sum_{\{s,t\}}\exp \left[ -%
\mathcal{H}(s,t)\right] .
\end{equation}%
and regarded as a partition function of a statistical model:
\begin{eqnarray}
&&\mathcal{H}=\sum_{\langle pq\rangle }\left[
J(s_{p}s_{q}+t_{p}t_{q})+J_{4}s_{p}s_{q}t_{p}t_{q}+J_{0}\right]  \notag \\
&&-\frac{i3\alpha }{2}\sum_{p}(s_{p}-t_{p})+\frac{i3\alpha }{4}\sum_{\langle
pqr\rangle }(s_{p}s_{q}s_{r}-t_{p}t_{q}t_{r}),  \label{SM}
\end{eqnarray}%
where $s_{p}$ and $t_{p}$ are the auxiliary Ising spins in the ket and the
bra layers, the 2-spin and 4-spin couplings are given by
\begin{equation}
J=\frac{1}{4}\log \frac{1+h^{2}-2h\sin \theta }{1+h^{2}+2h\sin \theta },%
\text{ }J_{4}=\frac{1}{4}\log \frac{4h^{2}\cos ^{2}\theta }{%
1+h^{4}+2h^{2}\cos 2\theta },  \notag
\end{equation}%
and the parameter $\alpha$ has been fixed as $e^{i6\alpha}=-1$. This is a two-dimensional triangular lattice Ashkin-Teller model with
\textit{imaginary} magnetic fields and \textit{imaginary} three-spin triangular
face interactions, which is derived from a topologically ordered wavefunction
for the first time. The imaginary terms of Eq. (\ref{SM}) originated from
the negative signs in the DS wavefunction, and this statistical model is
PT-symmetric, i.e., invariant under the combined operation of parity
symmetry P ($s_{p}\rightarrow -s_{p}$ and $t_{p}\rightarrow -t_{p}$) and
time-reversal symmetry T ($i\rightarrow -i$). This PT-symmetry ensures that
all eigenvalues of the transfer matrix operator of the partition function are either
real or complex conjugate pairs\cite{PTsymmetry}. Since this model has the
symmetry $h_{x}\rightarrow -h_{x}$, the tuning parameters are limited to $%
0\leq h\leq 1$ and $-\pi /2\leq \theta \leq $ $\pi /2$ in the following. In general, this model cannot be solved analytically except the following two special limits.

When $h_{x}=0$, we have $J_{4}\rightarrow \infty $, the Ising spins in two
layers are locked together, and the imaginary terms vanish, leading to an
inter-layer partial order $\langle st\rangle =\pm 1$. The norm of the DS and
TC wavefunctions become identical\cite{HuangWei1,XuZhang2018} and Eq. (\ref{SM}) is reduced to a two-dimensional triangular lattice Ising model: $%
\mathcal{H}=-2J\sum_{\langle pq\rangle }\tau_{p}\tau_{q}$ with $J=\frac{1}{2}\log \frac{
1-h_{z}}{1+h_{z}}$. For $h_{z}>0$, it is known that a ferromagnetic critical point\cite%
{Trangle_Ising} exists at $h_{z}=(3^{1/4}-1)/(3^{1/4}+1)$ and denoted by $H$. This critical point separates the intra-layer ferromagnetic phase and intra-layer
paramagnetic phase, corresponding to dilute loop phase and the DS phase,
respectively. For the antiferromagnetic coupling $h_{z}<0$, however, there
is no phase transition up to the multicritical point ($h_{z}=-1$) due to the
presence of spin frustration\cite{Anti_Ising_CFT}.

When $h=1$ and $-\pi/2<\theta <\pi /2$, the inter-layer coupling $J_{4}=0$
and the model of the decoupled layer is reduced to
\begin{equation}
\mathcal{H}_{s}=J\sum_{\langle pq\rangle }s_{p}s_{q}-\frac{i3\alpha }{2}%
\sum_{p}s_{p}+\frac{i3\alpha }{4}\sum_{\langle pqr\rangle }s_{p}s_{q}s_{r},
\label{O(-1)}
\end{equation}%
where $J=\frac{1}{4}\log \frac{1-\sin \theta }{1+\sin \theta }$. In the loop representation, the partition function of $\mathcal{H}_{s}$ becomes a single-layer $O(-1)$ loop model:
\begin{equation}
\mathcal{Z}_{s}=2\sum_{\{\mathcal{L\}}}e^{-2Jl_{\mathcal{L}}}(-1)^{\#%
\mathcal{L}},  \label{EqO(-1)}
\end{equation}%
where $l_{\mathcal{L}}$ is the combined length of the closed contractible
loops. By mapping to the antiferromagnetic Potts model, two exactly solvable
critical points were found and denoted by $A$ and $E$, respectively\cite{O(-1)1982, O(-1)1988,O(-1)1989}. The point $A$ at $e^{2J}=\sqrt{2-\sqrt{3}}$ is characterized by a non-unitary CFT with a central charge $c=-3/5$ or an effective central charge $c_{\ast}=3/5$. Meanwhile, the point $E$ at $e^{2J}=\sqrt{2+\sqrt{3}}$ corresponds to the fixed point of the dense loop phase, which is described by a non-unitary CFT with a central charge $c=-7$ or an effective central charge $c_{\ast}=1$ \cite{Ringel2016}. However, when the two-spin coupling $J$ is very small, the imaginary terms of Eq. (\ref{O(-1)}) become dominate, and there emerge zeros of the partition function, where the weights for different configurations in the partition function exactly cancel and the free energy is no longer continuous. In the following, we will employ the tensor-network numerical methods to find another edge-singularity point $D$ at $e^{2J}\approx 1.18$, and the zeros of the partition function are actually distributed in the region $\sqrt{2-\sqrt{3}}\leq e^{2J}<1.18$, forming a spontaneously PTSB phase.

\begin{figure}[tbp]
\includegraphics[width=0.9\textwidth]{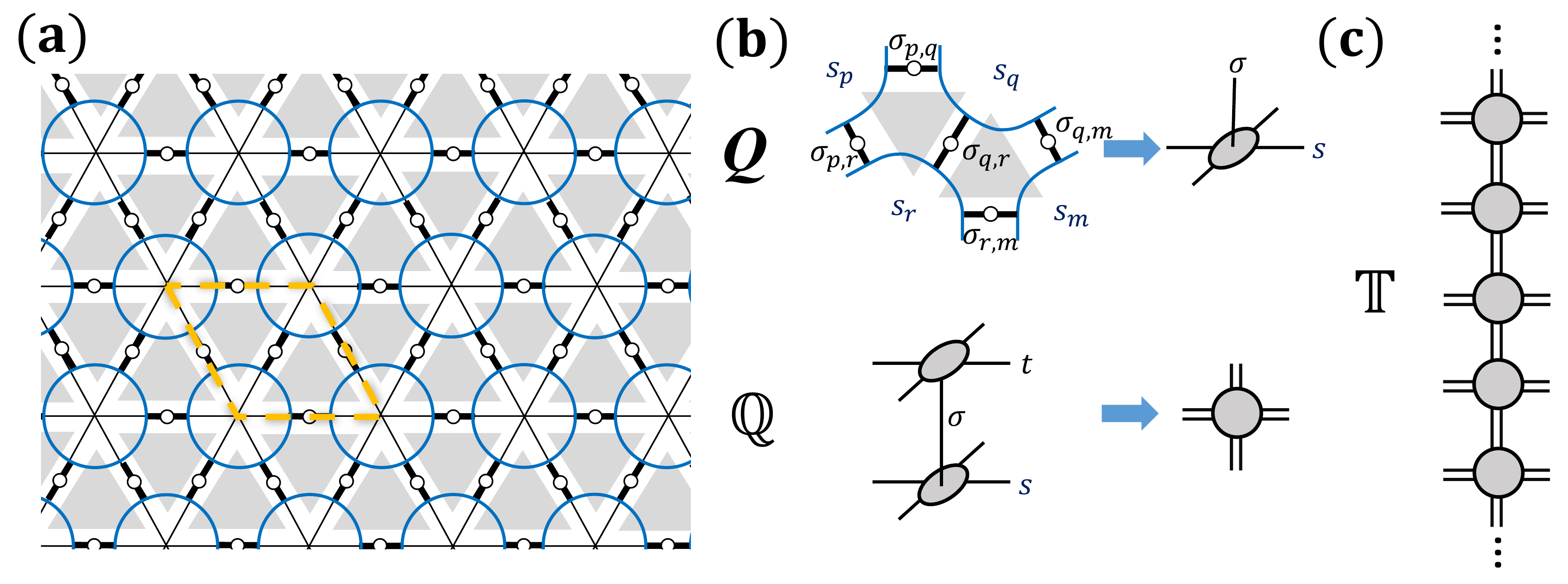}
\caption{\textbf{Local tensors and transfer matrix operator of the double tensor-network state.} (a) The TNS of the generic DS wavefunction. The large blue circles
denote the auxiliary spins $s_p$, small circles are the physical spins $%
\protect\sigma_{p,q}$, and the yellow dashed box shows the unit cell. (b)
The local double-line tensor $\bm{Q}$ on the square lattice is obtain by
contracting two local tensors on the hexagonal lattice. And $\mathbb{Q}$ is
the double-layer tensor. (c) The one-dimensional column-to-column transfer
matrix operator $\mathbb{T} $ of the double-layer tensor network.}
\label{TransferMatrix}
\end{figure}

\textbf{Tensor-network numerical calculations.} To establish the global
phase diagram of the generic DS wavefunction, we will apply the numerical
CTMRG method \cite{CTM_Nishino,CTM_Vidal,CTM_Corboz,fishman_faster_2018} to
the PT-symmetric statistical model Eq. (\ref{SM}) with negative Boltzmann
weights. In order to obtain the local tensors of TNS, each auxiliary spin $%
s_p$ is represented by a circle in Fig. \ref{TransferMatrix}a, and then the
generic wavefunction is expressed as a double-line TNS, whose local tensors
are defined on the vertices of the hexagonal lattice. For convenience, we transform
the hexagonal lattice into the square lattice by combining two local tensors
with nearest-neighbor triangle faces into one local tensor $\bm{Q}$.
Contracting the physical indices of $\bm{Q}$ and $\bm{Q}^*$ leads to the
local double-layer tensor $\mathbb{Q}$, as shown in Fig. \ref{TransferMatrix}%
b, and the wavefunction norm is expressed by
\begin{equation}  \label{double_layer__TNS}
\mathcal{Z}=\text{tTr}\left( \underset{\text{vertex}}{\bigotimes }\mathbb{Q}%
\right) =\text{Tr}(\mathbb{T}^{L_{x}}),
\end{equation}%
where \textquotedblleft tTr" denotes the tensor contraction over all
auxiliary indices, $\mathbb{T}$ is the column-to-column transfer matrix
operator displayed in Fig. \ref{TransferMatrix}c, and $L_{x}$ is the number
of columns. To determine the phase diagram, we need to calculate the dominant
eigenvalues of transfer operator $\mathbb{T}$, which contains the crucial
information about the phase transitions. Since the spectrum can be complex, we sort the eigenvalues according to their moduli.

\begin{figure}[tbp]
\centering	
\includegraphics[width=0.82\textwidth]{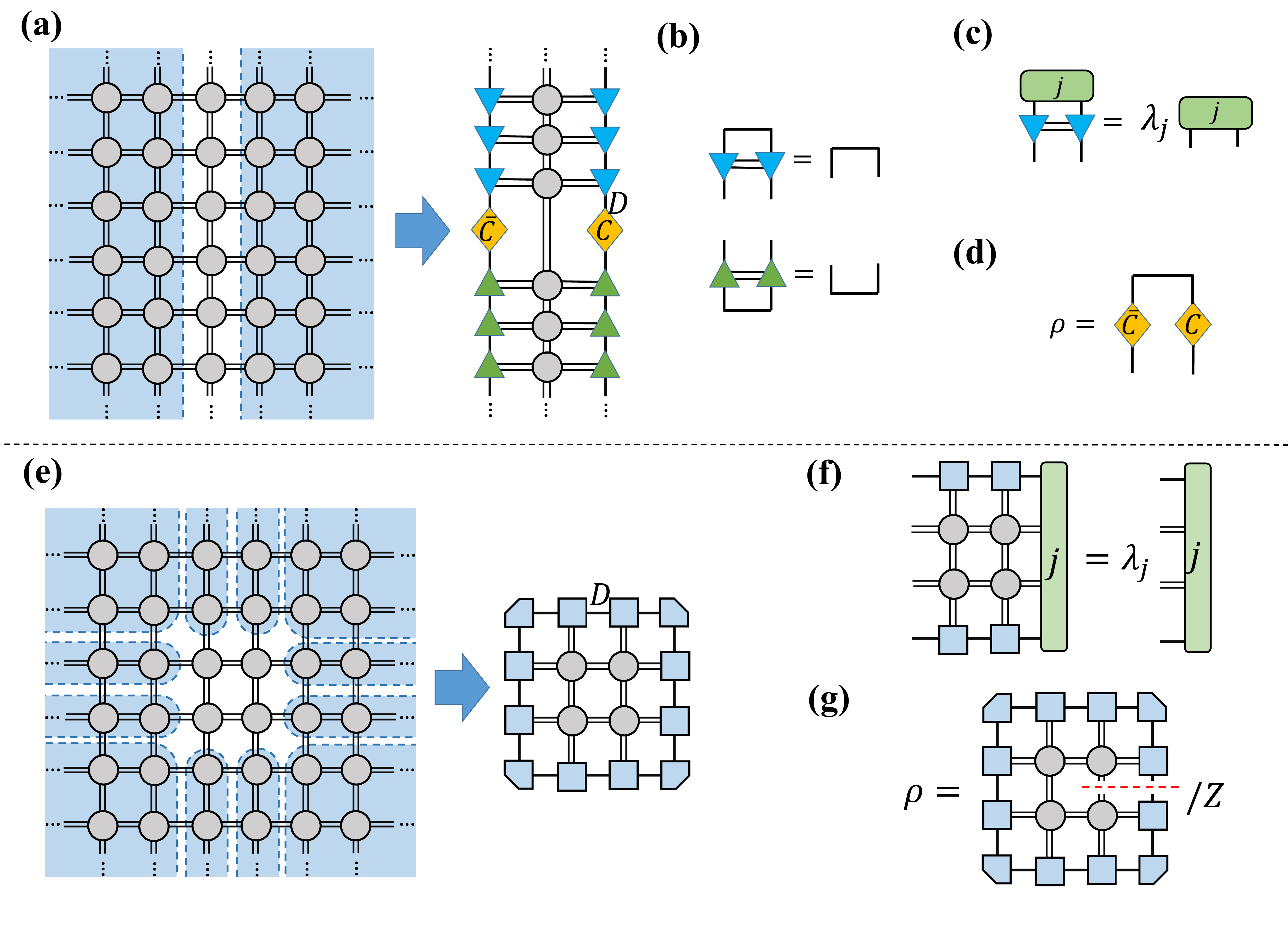}
\caption{\textbf{Main procedures of the tensor-network numerical methods.} (a) In the VUMPS algorithm, the environment of an infinite tensor network is approximated by the infinite matrix product state with a bond dimension $D$, and the MPS is optimized by maximizing the tensor network with variational method. The green and blue triangles are different isometric tensors, and the orange diamond is the central matrix $C$. (b) The isometric tensors satisfy the canonical conditions. (c) The eigen-equation of the column-to-column transfer operator $\mathbb{T}$. (d) The reduced density matrix $\protect\rho$ is expressed by the center matrices. (e) In the CTMRG method, the environment of an infinite tensor network is approximated by the edge fixed point tensors (squares) and corner fixed point tensors (snip single corner squares) with a bond dimension $D$. (f) The eigen-equation of the column-to-column transfer operator $\mathbb{T}$ is approximated by the edge tensors and local double tensors. (g) The reduced density matrix $\protect\rho$ is represented by the edge and corner tensors, where $Z$ is the normalization factor. }
\label{FigCTMRG}
\end{figure}

\begin{figure}[tbp]
\centering	
\includegraphics[width=0.95\textwidth]{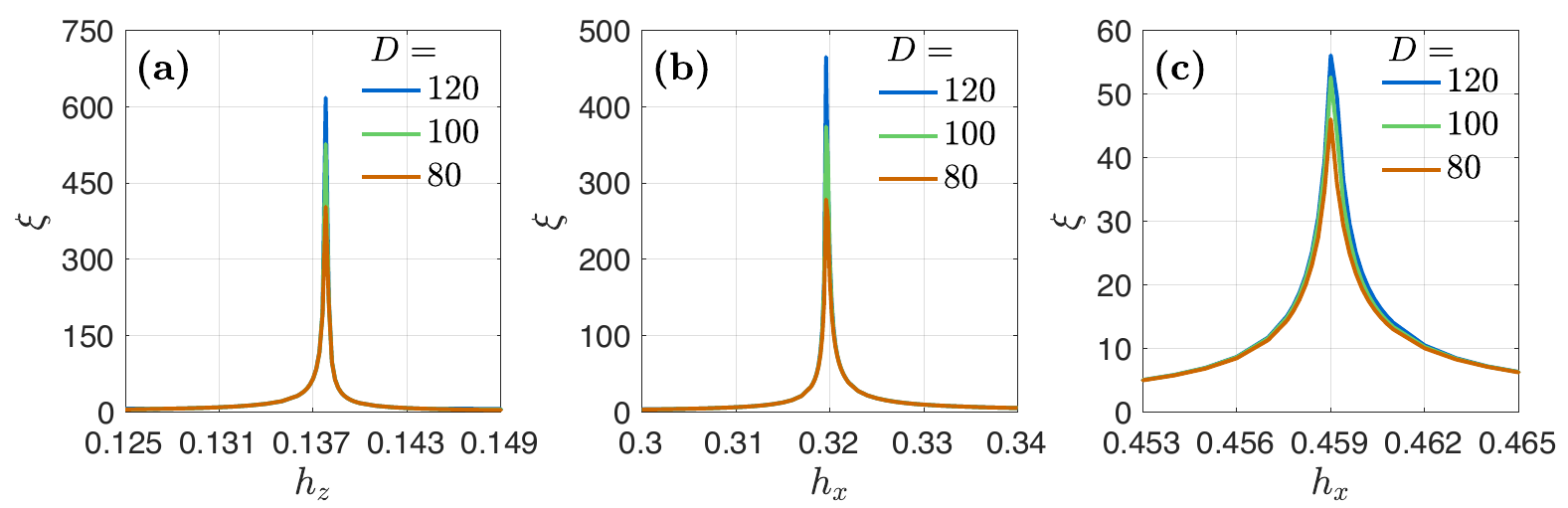}
\caption{\textbf{Correlation length of the generic TC wavefunction.} (a) The correlation length $\protect\xi$ with different bond dimensions $D$ along the $h_{x}=0.1$ cut. (b) Along the $h_{z}=-0.1$ cut. (c) Along the $h_{z}=0.7-h_{x}$ cut.}
\label{FigData4}
\end{figure}

As a comparison, we first calculate the phase diagram for the generic TC wavefunction. Since the imaginary terms are absent, Eq. (\ref{SM}) reduces to the two-dimensional Ashkin-Teller model on a triangular lattice, and the two-column transfer operator $\mathbb{T}\mathbb{T}^\text{t}$ is Hermitian. So the powerful variational uniform matrix-product-state (VUMPS) method\cite{VUMPSZauner,VUMPSFaster,VUMPSNote} can be applied, and its main procedures are summarized in Figs.\ref{FigCTMRG}a-\ref{FigCTMRG}d. In Fig. \ref{FigData4}a, we show the correlation length $\xi$ as a function of $h_{z}$ along the $h_{x}=0.1$ cut, and a sharp peak appears at $h_{z}=0.1378$, indicating a continuous phase transition from the inter-layer partial ordered phase ($\langle s\rangle=\langle t\rangle=0,\langle st\rangle\neq 0$) to the ferromagnetic phase ($\langle s\rangle\neq0,\langle t\rangle\neq0$). In Fig. \ref{FigData4}b, $\xi$ along the $h_{z}=-0.1$ cut is displayed, and a sharp peak around $h_{x}=0.3196$ represents a continuous phase transition from the partial ordered phase to the paramagnetic phase ($\langle s\rangle=\langle t\rangle=\langle st\rangle=0$). Moreover, in Fig. \ref{FigData4}c, a less sharper peak appears around $h_{z}=0.4590$ along the $h_{z}=0.7-h_{x}$ cut, implying the phase transition from the ferromagnetic phase to the
paramagnetic phase.

\begin{figure}[tbp]
\centering
\includegraphics[width=0.80\textwidth]{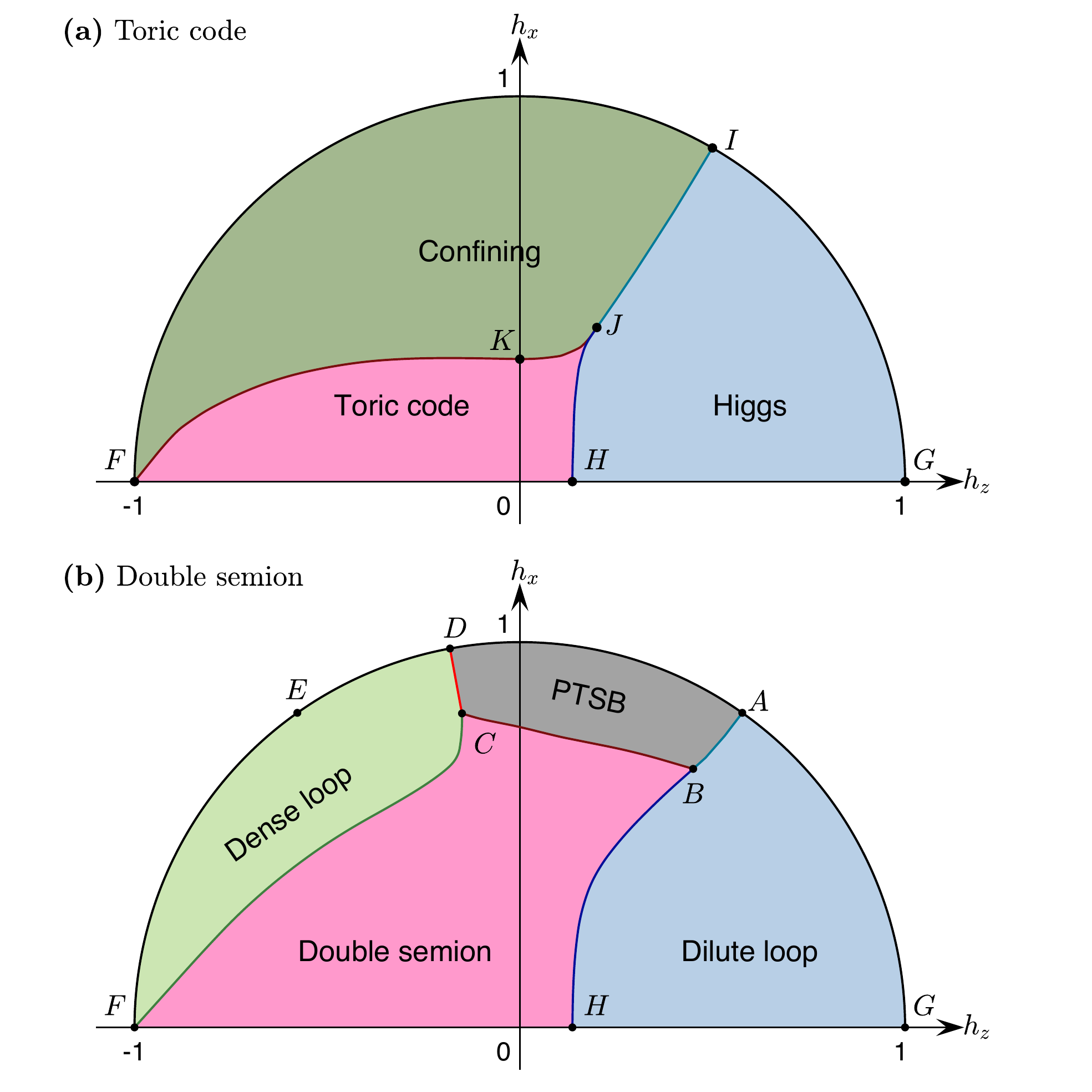}
\caption{\textbf{Global phase diagrams of the generic TC and DS wavefunctions.} (a) The phase diagram of the generic TC wavefunction. The Higgs phase and confining phase correspond to electric and magnetic anyon condensed phases, respectively. (b) The phase diagram of the generic DS wavefunction. Adjacent to the DS phase, there exist gapped dilute loop phase, gapless dense loop phase, and spontaneously PTSB phase. The phase transition lines $AB$, $BC$, and $CF$ consist of exceptional points, while the $CD$ line is a discontinuous phase transition.}
\label{FigPhaseDiagram}
\end{figure}

It has been proved that\cite{ZhuGuoYi_TC} the partial ordered phase, ferromagnetic phase and paramagnetic phase of the Ashkin-Teller model correspond to the TC phase, Higgs phase and confining phase, respectively. So a global phase diagram is obtained and displayed in Fig. \ref{FigPhaseDiagram}a. The TC phase is enclosed by the Higgs and confining phases, corresponding to the electric and magnetic anyon condensed phases, respectively. Both condensed phases are gapped, and there is a critical line $IJ$ between them, characterized by the Kosterlitz-Thouless transition. The points $I$ and $J$ are exactly determined as $(h_x,h_z)=(\sqrt{3}/2, 1/2)$ and $(h_x,h_z)=(2/5,1/5)$. Meanwhile, the critical lines $JH$ and $JKF$ belong to the two-dimensional Ising universality class. When $J=J_{4}$, we have $(h_{x}-1)^{2}+(h_{z}-1)^{2}=1$, on which the Ashkin-Teller model is mapped to the four-state Potts model, whose critical point is just the tricritical point $J$.

\begin{figure}[tbp]
\centering	
\includegraphics[width=0.85\textwidth]{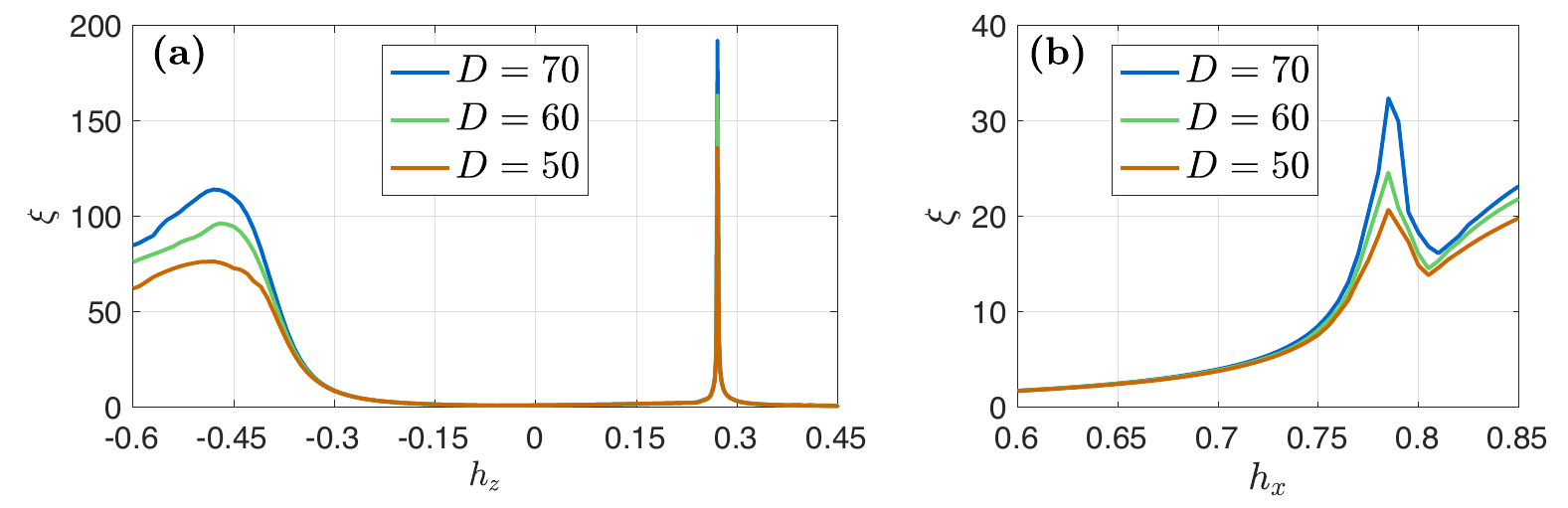}
\caption{\textbf{Correlation length of the generic DS wavefunction.} (a) The correlation length $\protect\xi $ with various bond dimensions along the $h_{x}=0.5$ cut. (b) Along
the $h_{x}$-axis.}
\label{FigData}
\end{figure}

\begin{figure}[tbp]
\centering
\includegraphics[width=0.95\textwidth]{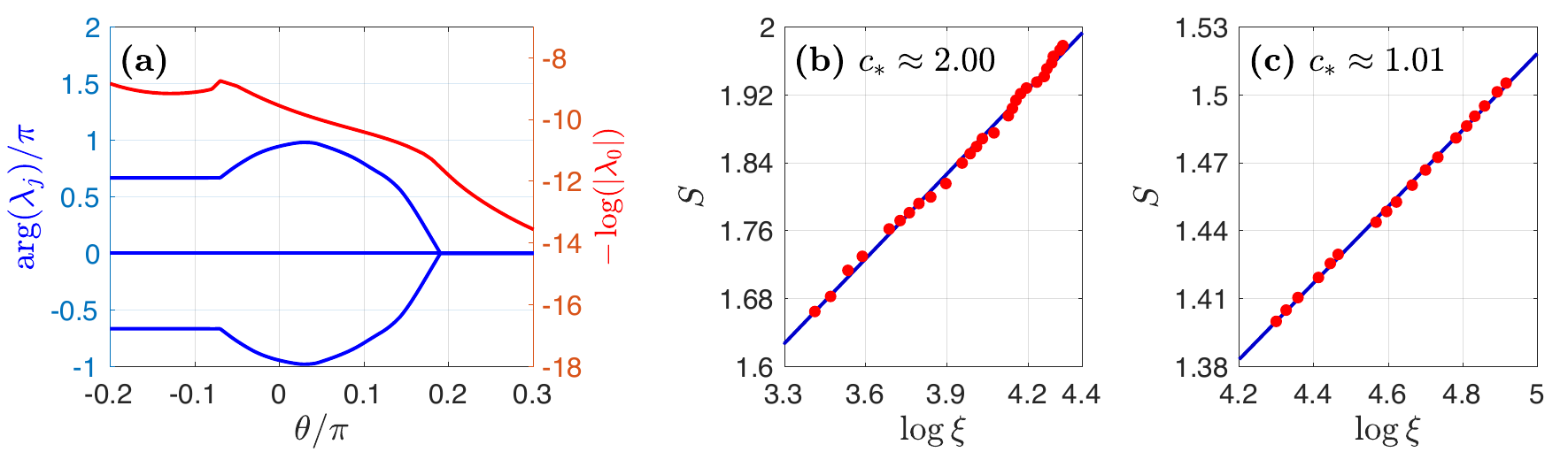}
\caption{\textbf{The leading eigenvalues and their arguments plus the scaling of central charges.} (a) The logarithm of the modulus of the leading eigenvalue $%
\protect\lambda_0$ (red line) and the arguments of three dominant
eigenvalues (blue lines) of the transfer matrix operator along the circle $%
h=0.85$. (b) The finite scaling of the entanglement entropy at the point $%
(h_{x},h_{z})=(0.5,-0.8)$ inside the dense loop phase, where $S$ is the
entanglement entropy. The blue line is $S=(c_{\ast }/6)\log \protect\xi %
+S_{0}$ with the estimated central charge $c_{\ast }\approx 2.00$. (c) At
the point $(h_x,h_z)=(0.500,-0.470)$ on the transition line $CF$ between the
DS phase and the dense loop phase, the estimated central charge is $c_*
\approx 1.01$.}
\label{FigArg}
\end{figure}

To establish the phase diagram of the generic DS wavefunction, we employ the
CTMRG method to calculate the correlation length $\xi$. The main ideas of
the CTMRG method are summarized in Figs. \ref{FigCTMRG}e-\ref{FigCTMRG}g, and
the numerical results are displayed in Fig. \ref{FigData} and Fig. \ref%
{FigArg}. In Fig. \ref{FigData}a, the correlation length is shown as a
function of $h_{z}$ along the cut $h_{x}=0.5$. As $h_{z}$ decreases from
unity, a sharp peak first appears at $h_{z}\approx 0.271$, indicating a
phase transition from the dilute loop phase to the DS phase. The peak
positions are nearly same as various bond dimensions. When $h_{z}$ is
further decreased, a hump appears around $h_{z}\approx -0.470$ and gradually
becomes a peak with increasing bond dimension. After this hump, the model
enters into the dense loop phase. By fitting the scaling relation of the entanglement entropy\cite{Finite_Entanglement_Scaling} at the point $(h_{x},h_{z})=(0.5,-0.8)$ inside the dense loop phase, we estimate an effective central charge $c_{\ast}\approx 2.00$ shown in Fig. \ref{FigArg}b. Considering that the exactly fixed point $E$ is contained this region and is described by the non-unitary CFT with a central charge $c=-7\times 2$, we then conclude that the gapless dense loop phase is characterized by the non-unitary CFT with an effective central charge $c_*=2$.

Moreover, the correlation length $\xi $ is also calculated along the $h_{x}$-axis and shown in Fig. \ref{FigData}b. As $h_{x}$ is increased, a peak appears around $h_{x}= 0.785$ and is enhanced by larger bond dimensions. For $h_{x}>0.785$, the model enters into a special phase, where the largest eigenvalue of the transfer matrix is not unique and the resulting phase \textit{spontaneously} breaks the PT-symmetry\cite{Bender1998,Ashida2017}. To clearly see the presence of the PTSB phase, several leading eigenvalues of the transfer matrix along the circle line $h=0.85$ are displayed in Fig. \ref{FigArg}a. For $\theta>0.19\pi$, the model is in the gapped dilute loop phase, in which the dominant eigenvalues of the transfer matrix are real and positive. At $\theta\approx0.19\pi$, the arguments of the leading eigenvalues indicate an exceptional point, where the corresponding eigenvectors coalesce. After this exceptional point, the arguments of a pair of complex dominate eigenvalues vary continuously in the PTSB phase. Moreover, the absolute value of the leading eigenvalue (red line) shows a cusp at $\theta\approx-0.07\pi$ between the PTSB phase and dense loop phase, indicating a discontinuous change of the free energy. For $\theta<-0.07\pi$, the dominate eigenvalue is real in the dense loop phase and the arguments of the second and third largest eigenvalues become $\pm2\pi/3$, which are related to the momenta carried by the corresponding eigenvectors.

With these numerical results, a global phase diagram of the generic DS
wavefunction can be fully established and shown in Fig. \ref{FigPhaseDiagram}b.
From our numerical results, the $CF$ line describes a continuous phase transition without the spontaneous symmetry breaking, while the $BC$ line is a second-order phase transition with spontaneous PT-symmetry breaking. However, both phase transitions are characterized by a non-unitary CFT with an effective central charge $c_{\ast }\approx 1.01$, and one of the fitting is displayed Fig. \ref{FigArg}c. Meanwhile, the phase transition line $AB$ is described by the same non-unitary CFT as the critical point $A$. More importantly, it should be mentioned that the continuous PT-symmetry breaking transitions are composed of the exceptional points, where the corresponding leading eigenvectors coalesce. On the other hand, zeros of the partition function are distributed along the $AD$ line, and the transition line $CD$ corresponds to discontinuous ones.

\textbf{DISCUSSION}

\begin{figure}[tbp]
\centering
\includegraphics[width=0.9\textwidth]{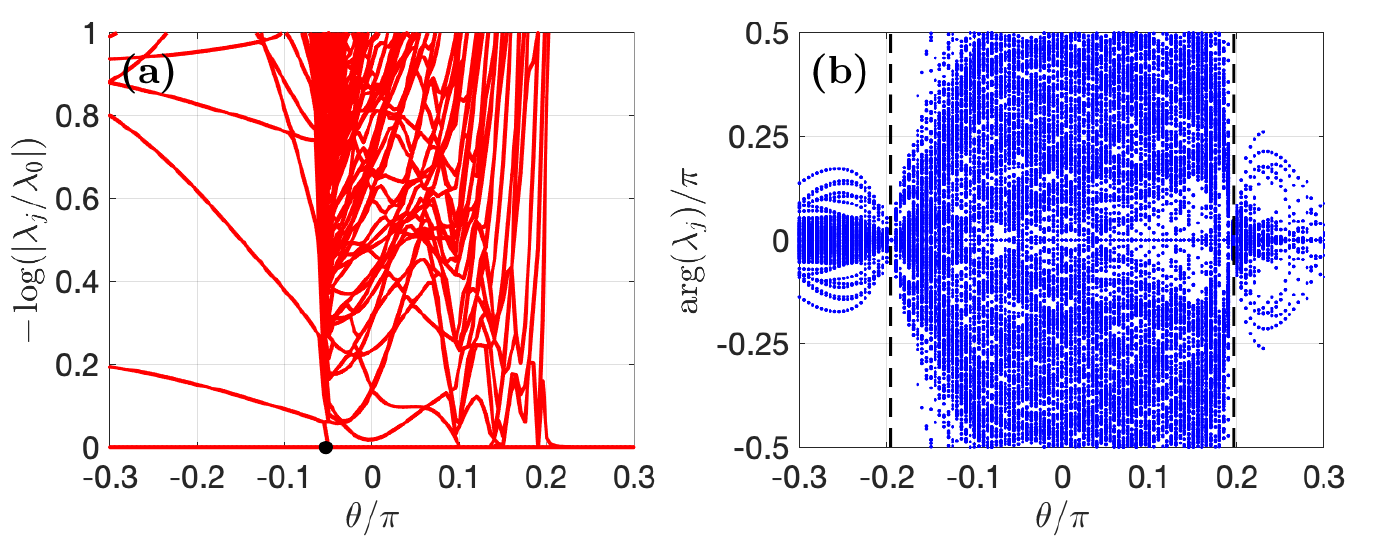}
\caption{\textbf{The spectrum of two-column transfer matrix $TT^{\text{t}}$ of the
single layer spin model.} With periodic boundary condition, we calculate the spectrum of $TT^{\text{t}}$ for the circumference $L_y=12$. (a) The logarithms of moduli of the
spectrum: $-\log(|\protect\lambda_j/\protect\lambda_0|)$, where $\protect\lambda_j$ is the $(j+1)$-th dominant eigenvalues. The black dot corresponds to the point $D$ in the phase diagram. (b) The arguments of the spectrum and the vertical dash lines corresponding to two exactly solvable points $A$ and $E$ of the phase diagram.}
\label{FigData-Spectrum}
\end{figure}

It should be emphasized that, three non-topological phases adjacent to the DS phase in Fig. \ref{FigPhaseDiagram}b are determined by two decoupled two-dimensional triangular lattice Ising-spin model Eq. (\ref{O(-1)}). In particular, the PTSB phase corresponds to the region with zeros of the partition function of the single-layer Ising-spin model or the $O(-1)$ loop model. Moreover, on the decoupled line $h=1$, the single-layer spin model allows us to perform exact diagonalization on a larger lattice size. For the two-column transfer operator $TT^{\text{t}}$ with periodic boundary condition, the
arguments of the eigenvalues related to the lattice momenta vanish. As shown in Fig. \ref{FigData-Spectrum}a, the absolute values of the eigenvalue spectrum reveal a discontinuous phase transition point $D$ at ($\theta \approx -0.05\pi $), where many eigenvalues of the transfer matrix crosses and the partition function changes sign. In Fig. \ref{FigData-Spectrum}b, the arguments of the eigenvalue spectrum have clearly demonstrated the presence of the exceptional points at $A$ and $E$, where the spectra become completely real. This is not surprising because the exactly solvable non-Hermitian models described by non-unitary CFTs always have the entire real spectra\cite{Ardonne2011}. Since the transfer matrix can not be transformed into a Hermitian operator, the nature of the critical point $A$ and $E$ have the non-Hermitian effects of intrinsic signs. Although both points are exceptional points, there is an essential difference: the dominant eigenvectors coalesce at the point $A$, while the coalescing eigenvectors at point $E$ are not. Meanwhile, the point $D$ as the zero of the partition function has an edge-singularity.

Compared to the phase diagram of the TC phase, new physics is introduced by the intrinsic signs in the DS wavefunction. Since the bosonic anyon excitations of the DS phase are related to the electric charge anyons of the TC phase, the gapped dilute loop phase corresponds to the Higgs phase, while the dense loop phase corresponds to the confining phase even though it is a gapless phase. But the PTSB phase is an emerging new phase, where the auxiliary Ising spins oscillate spatially and the pairs of the bosonic anyons condense. Moreover, there is a logarithmical attraction among the electric charge anyons along the transition line $IJ$ because of its description of a unitary CFT. Meanwhile, along the transition line $AB$, there are logarithmic repulsive interaction among the the pairs of bosonic anyons due to the negative scaling dimensions of the auxiliary spins from the non-unitary CFT.

So the quantum effects of the negative signs in a generic DS wavefunction have been explored using tensor network representation. The norm of this wavefunction is mapped to the partition function of a two-dimensional triangular lattice Ashkin-Teller model with
imaginary magnetic fields and imaginary three-spin triangular face interactions. Using the CTMRG method, the global phase diagram has been fully established. Adjacent to the DS topological phase, we found a gapped dilute loop phase, a gapless dense loop phase described by non-unitary CFT, and a PTSB phase with zeros of the partition function. Our study has thus proved that the negative signs in the topological wavefunction are intrinsic and deeply connected to negative Boltzmann weights of a PT-symmetric statistical model, shedding new light on the understanding of topologically ordered phases of matter.

\textbf{METHODS}

The phase diagrams in Fig. \ref{FigPhaseDiagram} are determined from
numerical tensor-network methods, and we will give their main ideas. The key
challenge in the numerical tensor-network method is the contraction of
tensor network generated by $\mathbb{Q}$. In general, the contraction of an
infinitely large tensor network is implemented by finding the approximate
environments with controllable errors. In our numerical calculation, we use
the infinite matrix product state (iMPS) or the corner transfer matrix (CTM)
to approximate the environments, as shown in the Figs. \ref{FigCTMRG}a and %
\ref{FigCTMRG}e.

The transfer matrix operator $\mathbb{T}\mathbb{T}^\text{t}$ for the norm of
the generic TC wavefunction is a Hermitian operator, so we can approximate
its dominant eigenvector using mixed canonical iMPS with bond dimension $D$
and optimize the iMPS using the VUMPS algorithm\cite%
{VUMPSNote,VUMPSZauner,VUMPSFaster}. The $(j+1)$-th dominant eigenvalues $%
\lambda_j$ of the row-to-row transfer operator and the reduced density
matrix $\rho$ can be calculated from the isometric tensors and the central
matrix $C$ of iMPS, as shown in Figs. \ref{FigCTMRG}c and \ref{FigCTMRG}d.
Then the correlation length $\xi$ and entanglement entropy $S$ can be
obtained: $\xi=-2/\log(|\lambda_1/\lambda_0|)$ and $S=-\text{Tr}%
(\rho\log\rho)$.

Since the transfer matrix operator $\mathbb{T}\mathbb{T}^\text{t}$ for the
norm of the generic DS wavefunction is non-Hermitian , its dominant
eigenvalues can not be worked out using variational method. Instead, we
solve the partition function using the CTM method, the environment is
approximated using edge tensors and corner tensors with tuning bond
dimension $D$. These tensors can be optimized using the CTMRG algorithm\cite%
{CTM_Nishino,CTM_Vidal,CTM_Corboz,fishman_faster_2018}. The correlation
length and entanglement entropy can be also calculated, where $\lambda_j$ is
defined in the Fig. \ref{FigCTMRG}f and $\rho$ is defined in Fig. \ref%
{FigCTMRG}g.

We should emphasis that the CTMRG calculations do not converge when all dominant eigenvalues are complex. However, due to our double layer construction, it can be derived that the most dominant eigenvalue of the double layer transfer operator is always real and positive. On the decoupled line, if the single layer transfer operator has a pair of dominant eigenvalues $|\lambda|e^{\pm i\phi}$, then the equal moduli dominant eigenvalues of the corresponding double layer transfer operator are $%
|\lambda|^2, |\lambda|^2e^{\pm i2\phi}$, where $|\lambda|^2$ is two-fold
degenerate. When the couplings between two layers are turned on, there are small splits between moduli of these dominant eigenvalues, and there is only one dominant one positive dominant eigenvalue strictly speaking. Thus the CTMRG calculations can slowly converge in the PTSB phase. When $h$ approach to $1$, the splits gradually vanish and the convergence of CTMRG algorithm is more and more slower. At $h=1$, there are several dominant eigenvalues with identical moduli, and the CTMRG algorithm is completely failed due to the intrinsic sign problem in the PTSB phase.

\bibliography{mybibtex}
\bibliographystyle{unsrt}

\textbf{Acknowledgments}

The authors are indebted to G. Y. Zhu for his earlier collaborations on this
project. The research is supported by the National Key Research and
Development Program of MOST of China (2016YFYA0300300 and 2017YFA0302902).

\textbf{Author Contributions}

{G.M.Zhang initiated and supervised this project, Q.Zhang conducted the
derivations and numerical calculations, W.T.Xu and Z.Q.Wang joined the
discussions, and GMZ wrote the paper with the helps of Q. Zhang and W.T.Xu.}

\textbf{Additional information}

\textbf{Competing interests}: The authors declare no competing financial interests.

\textbf{Materials and Correspondence}. Correspondence and requests for materials should be addressed to GMZ. (email:gmzhang@tsinghua.edu.cn) 

\end{document}